\documentclass{ws-ijmpa}

\begin{document}

\markboth{Roustam Zalaletdinov}
{The Averaging Problem in Cosmology and Macroscopic Gravity}

%
\catchline{}{}{}{}{}
%

\title{THE AVERAGING PROBLEM IN COSMOLOGY \\ AND MACROSCOPIC GRAVITY}

\author{ROUSTAM ZALALETDINOV}

\address{ICRANet Coordinating Center\\
P.le della Repubblica, 10, 65122 - Pescara (PE), Italy\\
zala@icra.it}

\maketitle

\begin{history}
\received{Day Month Year}
\revised{Day Month Year}
\end{history}

\begin{abstract}
The averaging problem in cosmology and
the approach of macroscopic gravity to resolve the
problem is discussed. The averaged Einstein equations of macroscopic gravity
are modified on cosmological scales by the macroscopic gravitational correlation
tensor terms as compared with the Einstein equations of general
relativity. This correlation tensor satisfies a system
of structure and field equations. An exact cosmological solution
to the macroscopic gravity equations for a constant macroscopic gravitational
connection correlation tensor for a flat spatially homogeneous,
isotropic macroscopic space-time is presented.
The correlation tensor term in the macroscopic Einstein equations has
been found to take the form of either a negative
or positive spatial curvature term. Thus, macroscopic gravity provides a cosmological
model for a flat spatially homogeneous, isotropic Universe
which obeys the dynamical law for either an open or closed Universe.

\end{abstract}

\ccode{PACS numbers: 04.20. q, 04.20.Cv, 98.80. k, 98.80.Hw}

\section{The Averaging Problem in Cosmology}

The real Universe is lumpy, not smooth. It appears to be
isotropic and homogeneous at the very large scales as supported by the observed
isotropy of the cosmic microwave background. The present state of the actual
Universe is, however, neither homogeneous nor isotropic on scales
$\lesssim 100$ \textit{Mps}. Since the Friedmann-Lema\^{\i}tre-Robertson-Walker
(FLRW) models, and more generally, relativistic cosmology of the Universe
relies on the Cosmological principle, the problem of dynamical justification
of the large-scale smooth structure, given inhomogeneous dynamics on
smaller scales, takes the central place in relativistic cosmology\cite
{Elli:1984}. This states the averaging problem in cosmology which calls for
an averaged description of the large-scale cosmological structure and its
evolution. Most of approaches to the averaging
problem to directly average out the Einstein equations and/or to describe
inhomogeneous cosmological models which are homogeneous and isotropic on
average made use of a perturbation theory and/or employed non-covariant
averaging schemes (see Refs.~{\refcite{Kras:1996}--\refcite{Para-Sing:2008}} for
review and references therein). The results obtained are
considered as having brought neither a rigorous formalism, nor a physically
adequate approach.

The standard approach in relativistic cosmology is based on two main assumptions.
$\langle 1 \rangle$ A discrete matter
distribution (on each physically distinct scale of stars, galaxies, clusters
of galaxies, etc.) of the real lumpy Universe with
the energy-momentum tensor $T_{\alpha \beta }^{\mathrm{(discrete)}}$
can be adequately approximated by a smoothed hydrodynamic
energy-momentum tensor,
\begin{equation}
T_{\alpha \beta }^{\mathrm{(hydro)}}=\langle T_{\alpha \beta }^{\mathrm{
(discrete)}}\rangle ,  \label{<T>}
\end{equation}
usually taken as perfect fluid. $\langle 2 \rangle$ Upon the averaging (\ref{<T>}) the structure of the
Einstein field operator is kept unchanged so that the Einstein equations hold again,
\begin{equation}
\overline{R}_{\alpha \beta }-\frac{1}{2}\overline{g}_{\alpha \beta }\overline{R}
=-\kappa T_{\alpha \beta }^{\mathrm{(hydro)}},  \label{EEst}
\end{equation}
where $\overline{g}_{\alpha \beta}$, $\overline{R}_{\alpha \beta} $ and
$\overline{R}=\overline{g}^{\mu \nu} \overline{R}_{\mu \nu}$ are the averaged
metric and Ricci tensors and curvature scalar, with the
corresponding equations of motion
\begin{equation}
T_{;\beta }^{\alpha \beta \mathrm{(hydro)}{}}=0.  \label{em-st}
\end{equation}
The physical and geometrical justification that
(\ref{<T>}) does imply (\ref{EEst}),
\begin{equation}
R_{\alpha \beta }-\frac{1}{2}g_{\alpha \beta
}R\rightarrow \langle R_{\alpha \beta }-\frac{1}{2}g_{\alpha \beta }R\rangle =
\overline{R}_{\alpha \beta }-\frac{1}{2}\overline{g}_{\alpha \beta }
\overline{R},
\label{ee->EE}
\end{equation}
remains the central issue in the standard approach. There exists a closely related
problem of the physical status of general relativity as a classical theory
of gravity and the range of physical scales where
the Einstein equations hold. It is likely to be a microscopic
theory\cite{Tava-Zala:1996} physically adequate for the description of
gravitational fields created by point-like discrete matter sources. These
fundamental questions arise in the applications of the Einstein equations in
cosmology and other physical settings with continuous, extended matter
distributions.

\section{The Averaging Problem and Macroscopic Gravity}

Macroscopic gravity is a geometric, non-perturbative
approach\cite{Zala:1992}\cdash\cite{Zala:2007} to resolve the averaging problem by its
reformulation as the problem of the macroscopic
description of classical gravitation. Classical physical phenomena are
known to possess two levels of description\cite{Lore:1916},\cite{deGr-Sutt:1972}.
The microscopic description is based on a discrete
matter model, while the macroscopic description is based on the continuous
matter model with a suitable averaging procedure employed to make a
transition between them. The Lorentz theory of electrons and
Maxwell's electrodynamics are well known as the
microscopic and macroscopic electromagnetic theories.

The problem of finding a macroscopic theory of gravity corresponding general relativity
as a microscopic theory has a number of fundamental difficulties. Three main
issues must be addressed here: (a) definition of covariant space-time volume or
statistical averages for the tensor fields on
pseudo-Riemannian space-times; (b) averaging out of the (pseudo)-Riemannian geometry of
space-time and the nonlinear structure of the Einstein field operator, and
definition of the macroscopic gravitational field correlations; (c)
construction of smoothed, continuously distributed models of
self-gravitating media. The Einstein equations
themselves have been shown to be insufficient to be consistently
averaged out, since they become just a definition of a correlation function.
Therefore a geometric treatment of the structure of the macroscopic (averaged)
space-time is necessary.

Macroscopic gravity provides a covariant, geometric and non-perturbative
approach to the averaging problem. It is a classical theory of gravity with a built-in
gravitational correlation scale, which is a generalization of general relativity.
In a cosmological setting
it serves as a theory of the large-scale gravitation. The theory of macroscopic
gravity incorporates a covariant space-time volume
averaging procedure for tensor fields on (pseudo)-Riemannian space-times with
well-defined analytic properties of average tensor fields. The structure
of the macroscopic (averaged) space-time geometry has been found by developing
a procedure of the averaging out of the Cartan structure equations
of the pseudo-Riemannian geometry of (microscopic) space-time, and the definitions
of the macroscopic gravitational correlation tensor fields have been given.
The problem of construction of models of continuously distributed self-gravitating media
has been also approached\cite{MRZ:2000},\cite{MRZ:2003}.

The macroscopic space-time geometry is a non-trivial generalization of the
metric affine connection geometries, which involves, as a part, the Riemannian and
non-Riemannian geometric structures. The system of the field equations of
macroscopic gravity includes the macroscopic (averaged) Einstein
equations. There is a system of equations for the macroscopic gravitational
metric and connection correlation tensors. The structure of the  macroscopic gravity
equations enables one to answer
the fundamental question of relativistic cosmology about the physical
meaning and the range of applicability of the Einstein equations with a
continuous (smoothed) matter source. The macroscopic (averaged) Einstein
equations of macroscopic gravity become the Einstein equations of general
relativity,
\begin{equation}
M_{\alpha \beta }-\frac{1}{2}G_{\alpha \beta }G^{\mu \nu }M_{\mu \nu
}=-\kappa T_{\alpha \beta }^{\mathrm{(hydro)}},  \label{MG->EE}
\end{equation}%
for the macroscopic metric tensor $G_{\alpha \beta }$ with a smoothed
energy-momentum tensor \emph{only if all macroscopic gravitational metric and
connection correlations vanish}.

\section{The System of Macroscopic Gravity Equations}

The system of the macroscopic gravity
equations and a procedure of its solving for the simplest case with no metric correlations, one connection
correlation tensor and the averaged matter energy-momentum tensor taken as
perfect fluid read as follows.

(I-A) Define a line element for the macroscopic space-time,
\begin{equation}
ds^{2}=G_{\alpha \beta }dx^{\alpha }dx^{\beta },
\label{@ds^2}
\end{equation}
where $G_{\alpha \beta }$ is the macroscopic metric tensor and
calculate the macroscopic Levi-Civita connection coefficients $\overline{%
\mathcal{F}}^{\alpha }{}_{\beta \gamma }\equiv \langle \mathcal{F}^{\alpha
}{}_{\beta \gamma }{}\rangle$,
\begin{equation}
\overline{\mathcal{F}}^{\alpha }{}_{\beta \gamma }=\frac{1}{2}G^{\alpha
\epsilon }(G_{\beta \epsilon ,\gamma }+G_{\gamma \epsilon ,\beta }-G_{\beta
\gamma ,\epsilon }),  \label{@Fb=dG}
\end{equation}
and the macroscopic Riemann curvature tensor $M^{\alpha }{}_{\beta \rho \sigma }$,
\begin{equation}
M^{\alpha }{}_{\beta \gamma \delta }=2\overline{\mathcal{F}}^{\alpha
}{}_{\beta \lbrack \delta ,\gamma ]}+2\overline{\mathcal{F}}^{\alpha
}{}_{\epsilon \lbrack \gamma }\overline{\mathcal{F}}^{\epsilon }{}_{%
\underline{\beta }\delta ]},  \label{@M=DFb}
\end{equation}
in terms of unknown macroscopic metric functions $G_{\alpha \beta }$.

(I-B) Provided the macroscopic space-time is highly symmetric, take the
averaged metric tensor $\bar{g}_{\alpha \beta }$ and averaged inverse metric tensor
$\bar{g}^{\alpha \beta }$ as
\begin{equation}
\bar{g}_{\alpha \beta }=G_{\alpha \beta },\quad \bar{g}^{\alpha \beta
}=G^{\alpha \beta },  \label{@g(-1)b=G(-1)}
\end{equation}
that means that there are no metric correlations in such cases.

(II-A) Determine the full set of linearly independent components of the
connection correlation tensor $Z^{\alpha }{}_{\beta \gamma }{}^{\mu}{}_{\nu \sigma }$,
\begin{equation}
Z^{\alpha }{}_{\beta \gamma }{}^{\mu }{}_{\nu \sigma }\equiv Z^{\alpha
}{}_{\beta \lbrack \gamma }{}^{\mu }{}_{\underline{\nu }\sigma ]}=\langle
\mathcal{F}^{\alpha }{}_{\beta \lbrack \gamma }{}\mathcal{F}^{\mu }{}_{%
\underline{\nu }\sigma ]}\rangle -\langle \mathcal{F}^{\alpha }{}_{\beta
\lbrack \gamma }{}\rangle \langle \mathcal{F}^{\mu }{}_{\underline{\nu }%
\sigma ]}\rangle ,  \label{Z6}
\end{equation}%
which {satisfies the algebraic conditions:

(i) the antisymmetry in the third and sixth indices,
\begin{equation}
Z^{\alpha }{}_{\beta \gamma }{}^{\mu }{}_{\nu \sigma }=-Z^{\alpha }{}_{\beta
\sigma }{}^{\mu }{}_{\nu \gamma },  \label{Z+Z=0}
\end{equation}

(ii) the antisymmetry in interchange of the index pairs,
\begin{equation}
Z^{\alpha }{}_{\beta \gamma }{}^{\mu }{}_{\nu \sigma }=-Z^{\mu }{}_{\nu
\gamma }{}^{\alpha }{}_{\beta \sigma },  \label{Z2+Z2=0}
\end{equation}

(iii) the algebraic cyclic identities,
\begin{equation}
Z^{\alpha }{}_{\beta \lbrack \gamma }{}^{\mu }{}_{\nu \sigma ]}=0,
\label{Z^dx=0}
\end{equation}

(iv) the equi-affinity property,
\begin{equation}
Z^{\epsilon }{}_{\epsilon \gamma }{}^{\mu }{}_{\nu \sigma }=0,  \label{TrZ=0}
\end{equation}%
and the algebraic symmetry of interchange of the index triples
from (\ref{Z+Z=0}) and (\ref{Z2+Z2=0}),
\begin{equation}
Z^{\alpha }{}_{\beta \gamma }{}^{\mu }{}_{\nu \sigma }=Z^{\mu }{}_{\nu
\sigma }{}^{\alpha }{}_{\beta \gamma }.  \label{Z3-Z3=0}
\end{equation}
There are generally 396 linearly independent components which represent
the set of unknown connection correlation functions.

(II-B) Require geometrical or physical conditions on the structure of the
connection correlation tensor $Z^{\alpha }{}_{\beta \gamma }{}^{\mu}{}_{\nu \sigma }$.
The simplest geometrical conditions are requirements of its being either
constant, or covariantly constant with respect to the {macroscopic
Levi-Civita connection, or Lie-constant with the respect to the isometry
group of the macroscopic metric tensor, and other conditions. Solve the
system of the equations on the connection correlation tensor due to
the conditions adopted.

(II-C) Solve the integrability conditions for the differential equations
(\ref{@DZ=0}) for the connection correlation tensor $Z^{\alpha }{}_{\beta
\gamma }{}^{\mu }{}_{\nu \sigma }$,
\begin{equation}
Z^{\epsilon }{}_{\beta \lbrack \gamma }{}^{\mu }{}_{\underline{\nu }\sigma
}M^{\alpha }{}_{\underline{\epsilon }\lambda \rho ]}-Z^{\alpha }{}_{\epsilon
\lbrack \gamma }{}^{\mu }{}_{\underline{\nu }\sigma }M^{\epsilon }{}_{%
\underline{\beta }\lambda \rho ]}+Z^{\alpha }{}_{\beta \lbrack \gamma
}{}^{\epsilon }{}_{\underline{\nu }\sigma }M^{\mu }{}_{\underline{\epsilon }%
\lambda \rho ]}-Z^{\alpha }{}_{\beta \lbrack \gamma }{}^{\mu }{}_{\underline{%
\epsilon }\sigma }M^{\epsilon }{}_{\underline{\nu }\lambda \rho ]}=0,
\label{@ic:DZ=0}
\end{equation}%
to find a set of unknown connection correlation functions.

(II-D) Solve the system of differential equations for the connection
correlation tensor }$Z^{\alpha }{}_{\beta \gamma }{}^{\mu }{}_{\nu \sigma }$,
\begin{equation}
Z^{\alpha }{}_{\beta \lbrack \gamma }{}^{\mu }{}_{\underline{\nu }\sigma
\parallel \lambda ]}=0,  \label{@DZ=0}
\end{equation}
where $\parallel $ is the covariant derivative with respect to
$\overline{\mathcal{F}}^{\alpha }{}_{\beta \gamma }$, to find a set of some
unknown connection correlation functions.

(II-E) Solve the quadratic algebraic conditions for the connection
correlation tensor $Z^{\alpha }{}_{\beta \gamma }{}^{\mu }{}_{\nu \sigma }$,
\begin{gather}
Z^{\delta }{}_{\beta \lbrack \gamma }{}^{\theta }{}_{\underline{\kappa }\pi
}Z^{\alpha }{}_{\underline{\delta }\epsilon }{}^{\mu }{}_{\underline{\nu }%
\sigma ]}+Z^{\delta }{}_{\beta \lbrack \gamma }{}^{\mu }{}_{\underline{\nu }%
\sigma }Z^{\theta }{}_{\underline{\kappa }\pi }{}^{\alpha }{}_{\underline{%
\delta }\epsilon ]}+Z^{\alpha }{}_{\beta \lbrack \gamma }{}^{\delta }{}_{%
\underline{\nu }\sigma }Z^{\mu }{}_{\underline{\delta }\epsilon }{}^{\theta
}{}_{\underline{\kappa }\pi ]}+  \nonumber \\
\quad Z^{\alpha }{}_{\beta \lbrack \gamma }{}^{\mu }{}_{\underline{\delta }%
\epsilon }Z^{\theta }{}_{\underline{\kappa }\pi }{}^{\delta }{}_{\underline{%
\nu }\sigma ]}+Z^{\alpha }{}_{\beta \lbrack \gamma }{}^{\theta }{}_{%
\underline{\delta }\epsilon }Z^{\mu }{}_{\underline{\nu }\sigma }{}^{\delta
}{}_{\underline{\kappa }\pi ]}+Z^{\alpha }{}_{\beta \lbrack \gamma
}{}^{\delta }{}_{\underline{\kappa }\pi }Z^{\theta }{}_{\underline{\delta }%
\epsilon }{}^{\mu }{}_{\underline{\nu }\sigma ]}=0,  \label{@ZZ=0}
\end{gather}
to find a set of unknown connection correlation functions.

(III-A) Determine the macroscopic non-Riemannian curvature tensor
$R^{\alpha }{}_{\beta \rho \sigma }$ from the macroscopic Riemann curvature
tensor $M^{\alpha }{}_{\beta \rho \sigma }$ and the connection
correlation tensor $Q^{\alpha }{}_{\beta \gamma \lambda }=
2Z^{\alpha}{}_{\varepsilon \gamma }{}^{\varepsilon }{}_{\beta \lambda }$,
\begin{equation}
R^{\alpha }{}_{\beta \rho \sigma }=M^{\alpha }{}_{\beta \rho \sigma}
+Q^{\alpha }{}_{\beta \rho \sigma }.  \label{@R=M+Q}
\end{equation}

(III-B) Define the affine deformation tensor $A^{\alpha }{}_{\beta \gamma}$,
\begin{equation}
A^{\alpha }{}_{\beta \gamma }=\overline{\mathcal{F}}^{\alpha }{}_{\beta
\gamma }-\Pi ^{\alpha }{}_{\beta \gamma },  \label{A}
\end{equation}
where a symmetric affine connection $\Pi ^{\alpha }{}_{\beta \gamma }$ for the
macroscopic non-Riemannian }curvature tensor $R^{\alpha }{}_{\beta \rho\sigma }$
is not compatible with the {macroscopic metric tensor }$G_{\alpha
\beta }$. There are {generally 40 unknown affine deformation functions.

(III-C) Require geometrical or physical conditions on the structure of the
affine deformation tensor $A^{\alpha }{}_{\beta \gamma }$. The simplest
geometrical conditions are requirements of its being
either constant, or covariantly constant with respect to the
macroscopic Levi-Civita connection, or Lie-constant with the respect to the
isometry group of the macroscopic metric tensor, and other conditions.
Solve the system of the equations on the affine deformation tensor due to
the conditions adopted.

(III-D) Solve the system of equations for the affine deformation tensor
$A^{\alpha }{}_{\beta \rho }$,
\begin{equation}
A^{\alpha }{}_{\beta \lbrack \sigma \parallel \rho ]}-A^{\alpha
}{}_{\epsilon \lbrack \rho }A^{\epsilon }{}_{\underline{\beta }\sigma ]}=-%
\frac{1}{2}Q^{\alpha }{}_{\beta \rho \sigma },  \label{@DA=Q}
\end{equation}%
to find a set of unknown affine deformation functions.

(IV-A) Assume that the averaged matter energy-momentum tensor
$\langle t_{\beta }^{\alpha \mathrm{(micro)}}\rangle $ can be taken as a
perfect fluid energy-momentum tensor,
\begin{equation}
\langle \mathbf{t}_{\beta }^{\alpha \mathrm{(micro)}}\rangle =
(\rho+p)u^{\alpha }u_{\beta }+p\delta _{\beta }^{\alpha },  \label{@<t>=pf}
\end{equation}
with the mass density $\rho $, the pressure }$p$, the fluid
4-velocity $u^{\alpha }$, $u^{\beta }u_{\beta }=-1$, and an equation
of state $p=p(\rho)$.

(IV-B) Solve the macroscopic gravitational correlation field equations
for the gravitational correlation energy-momentum tensor
$T_{\beta }^{\alpha {\text{(\emph{grav}) }}}${of macroscopic gravity,
\begin{equation}
(Z^{\alpha }{}_{\mu \nu \beta }-\frac{1}{2}\delta _{\beta }^{\alpha }Q_{\mu
\nu })G^{\mu \nu }=-\kappa T_{\beta }^{\alpha {(\text{\emph{grav})}}},
\label{@RicZ=T(grav)}
\end{equation}
where $Z^{\alpha }{}_{\mu \nu \beta }=2Z^{\alpha }{}_{\mu \varepsilon}{}^{\varepsilon }{}_{\nu \beta }$
and $Q_{\mu \nu }$ $=Z^{\varepsilon}{}_{\mu \nu \varepsilon }$, in terms of unknown macroscopic metric and
connection correlation functions. The energy-momentum of the macroscopic
gravitational correlation field }is conserved separately due to {(\ref{@DZ=0}).
Additional assumptions regarding the structure of the gravitational
correlation energy-momentum tensor $T_{\beta }^{\alpha {\text{(\emph{grav})}}}$ may be necessary here.

(IV-C) Solve the field equations for the macroscopic non-Riemannian
curvature tensor, $R^{\alpha }{}_{\beta [ \rho \sigma \parallel \lambda ] }=0$,
which can be written as
\begin{equation}
A^{\epsilon }{}_{\beta \lbrack \rho }R^{\alpha }{}_{\underline{\epsilon }%
\sigma \lambda ]}-A^{\alpha }{}_{\epsilon \lbrack \rho }R^{\epsilon }{}_{%
\underline{\beta }\sigma \lambda ]}=0,  \label{@AR-AR=0}
\end{equation}
to find a set of unknown affine deformation functions.

(IV-D) Solve the macroscopic Einstein equations,
\begin{equation}
G^{\alpha \epsilon }M_{\epsilon \beta }-\frac{1}{2}\delta _{\beta }^{\alpha
}G^{\mu \nu }M_{\mu \nu }=-\kappa \langle \mathbf{t}_{\beta }^{\alpha
\mathrm{(micro)}}\rangle -\kappa T_{\beta }^{\alpha {(\text{\emph{grav})}}},
\label{@M=<t>+Z}
\end{equation}
for the unknown metric functions.

The system of the macroscopic gravity equations (\ref{@ic:DZ=0})-(\ref{@ZZ=0}),
(\ref{@DA=Q}) and (\ref{@RicZ=T(grav)})-(\ref{@M=<t>+Z}) for the
unknown macroscopic metric, connection correlation and affine deformation
functions is a coupled system of the first and second order nonlinear
partial differential equations.

\section{An Exact Cosmological Solution}

Consider a flat spatially homogeneous, isotropic macroscopic space-time
given by the Robertson-Walker line element with cosmological time $t$,
\begin{equation}
ds^{2}=-dt^{2}+a(t)(dx^{2}+dy^{2}+dz^{2}),  \label{rw-flat}
\end{equation}
where $a(t)$ is the cosmological scale factor.

The simplest assumption on the structure of the connection correlation
tensor $Z^{\alpha }{}_{\beta \gamma }{}^{\mu }{}_{\nu \sigma }$ compatible
with the structure of macroscopic space-time (\ref{rw-flat}) is to require
\begin{equation}
Z^{\alpha }{}_{\beta \gamma }{}^{\mu }{}_{\nu \sigma }=\mathrm{const}
\label{Z6=const}
\end{equation}
that means from the physical point of view that the macroscopic
gravitational connection correlations do not change in time and space. An exact
solution\cite{CPZ:2005},\cite{Zala-vdHo:2008} to the macroscopic gravity equations has only
one non-trivial linearly independent component $12Z^{3}{}_{22}{}^{3}{}_{33}=%
\varepsilon $ of the connection correlation tensor
$Z^{\alpha }{}_{\beta\gamma }{}^{\mu }{}_{\nu \sigma }$ determined by an integration constant
$\varepsilon$. The gravitational correlation energy-momentum tensor (\ref{@RicZ=T(grav)}) has been
shown to have the structure
\begin{equation}
\kappa T_{\alpha \beta }^{{(\text{\emph{grav}})}}{}=\kappa G_{\alpha
\epsilon }T_{\beta }^{\epsilon {(\text{\emph{grav}})}}{}=\begin{bmatrix}
\varepsilon /a^{2} & 0 & 0 & 0 \\ 0 & -\varepsilon /3 a^{2} & 0 & 0 \\ 0 & 0 &
-\varepsilon /3 a^{2} & 0 \\ 0 & 0 & 0 & -\varepsilon /3 a^{2}
\end{bmatrix}  \label{Tgrav[2,2,3]real-cov}
\end{equation}
where $\rho _{\text{\emph{grav}}}=\varepsilon /\kappa a^{2}$ is the
energy density and $p_{\text{\emph{grav}}}=-\varepsilon /3\kappa a^{2}$
is the isotropic pressure of the macroscopic gravitational
correlation field. Thus, the macroscopic gravity equations fix the equation
of state for the macroscopic gravitational correlation field as
\begin{equation}
p_{\text{\emph{grav}}}=p_{\text{\emph{grav}}}(\rho _{\text{\emph{grav}}})
=-\frac{1}{3}\rho _{\text{\emph{grav}}}=-\frac{\varepsilon }{3a^{2}}.
\label{MG-EqSt}
\end{equation}
{The trace of }$T_{\beta }^{\alpha {\text{(\emph{grav}})}}$ is
\begin{equation}
T_{\epsilon }^{\epsilon {\text{(\emph{grav}})}}{}=
-\frac{2\varepsilon }{\kappa a^{2}}=-2\rho _{\text{\emph{grav}}},\quad
T_{\epsilon }^{\epsilon {\text{(\emph{grav}})}}<0~\mathrm{if~}\rho _{\text{\emph{grav}}}>0,
\label{Tr_Tgrav[2,2,3]real}
\end{equation}
that means from the physical point of view that the positive energy density
of the macroscopic gravitational correlation field, $\rho _{\text{\emph{grav}}}>0$,
corresponds to the binding energy of the Universe. Such
gravitational correlation field acts like tension in an elastic medium to
keep it intact. An amount of work should be performed and an amount of
energy should be spent to change its state, size and shape. A negative
pressure has the similar physical meaning and effect. The binding energy
density is decreasing with an increasing scale factor.

With the macroscopic distribution of the cosmological matter taken as the
perfect fluid energy-momentum tensor (\ref{@<t>=pf}), the macroscopic
Einstein equations (\ref{@M=<t>+Z}) read
\begin{equation}
\left( \frac{\dot{a}}{a}\right) ^{2}=\frac{\kappa \rho }{3}
+\frac{\varepsilon }{3a^{2}},\quad \frac{\ddot{a}}{a}=-\frac{\kappa }{6}(\rho +3p),
\label{macro-law}
\end{equation}
with the equations of state $p=p(\rho )$ of the averaged matter and (\ref{MG-EqSt}) for the
gravitational connection correlation field $Z^{\alpha }{}_{\beta \gamma}{}^{\mu }{}_{\nu \sigma }$.
The equations (\ref{macro-law}) look similar
to the Einstein equations of general relativity for either a closed or open
spatially homogeneous, isotropic FLRW space-time. However, they do have different
mathematical and physical, and therefore, cosmological content since,
firstly, the spatially homogeneous, isotropic cosmological macroscopic
space-time is flat (\ref{rw-flat}), and, secondly,
\begin{equation}
\kappa \rho _{\text{\emph{grav}}}=\frac{\varepsilon }{a^{2}}\neq -\frac{3k}{%
a^{2}},  \label{sp-curv-not-k}
\end{equation}
that is, $\varepsilon \neq -3k$ in general. Only if one
requires $12Z^{3}{}_{22}{}^{3}{}_{33}=\varepsilon =-3k$, the macroscopic
Einstein equations become exactly the Einstein equations of
general relativity for either a closed or open spatially homogeneous,
isotropic space-time, but, the macroscopic space-time has the flat spatially
homogeneous, isotropic geometry.

Thus, the theory of macroscopic gravity predicts that the macroscopic Einstein
equations for a flat spatially homogeneous, isotropic macroscopic space-time
with the constant macroscopic gravitational connection correlations
(\ref{Z6=const}) have the correlation term of the form of a spatial curvature
term. This is a \emph{dark spatial curvature} of the macroscopic gravitational correlation
filed since due to the field equations (\ref{@RicZ=T(grav)})-(\ref{@M=<t>+Z}) of macroscopic gravity
(i) it interacts only gravitationally with the macroscopic
gravitational field, (ii) it does not interact directly with the
energy-momentum tensor of averaged matter, (iii) it exhibits a negative
gravitational correlation pressure $p_{\text{\emph{grav}}}=-\rho _{\text{\emph{grav}}}/3$
when $\rho _{\text{\emph{grav}}}>0$. It is of fundamental
geometrical and cosmological significance that like the spatial curvature
term in the Einstein equation of general relativity, such a gravitational correlation term
in the macroscopic Einstein equations (\ref{macro-law}) of macroscopic gravity does not
contribute into either acceleration or deceleration of the Universe
in accordance with the equation of state (\ref{MG-EqSt}).

This exact solution of the macroscopic gravity equations reveals a very
non-trivial phenomenon from the point of view of the general-relativistic
cosmology: the macroscopic (averaged) large-scale cosmological evolution in
a flat Universe is governed by the dynamical evolution equations for either
a closed or open Universe depending on the sign of the energy
density $\rho _{\text{\emph{grav}}}$ of the macroscopic gravitational
correlation field in the dark spatial curvature term $\kappa \rho _{\text{\emph{grav}}}/3$.
Such a cosmological model shows that macroscopic
gravity is capable of providing a framework for a new cosmological paradigm
to reconsider the standard cosmological interpretation and treatment of the
observational data. Indeed, this macroscopic cosmological model has the
pseudo-Riemannian geometry of a flat spatially homogeneous, isotropic space-time. Therefore,
all measurements and observations are to be considered and designed for this
geometry. The dynamical interpretation of the observational data should be
considered and treated for the cosmological evolution of either a closed or
open spatially homogeneous, isotropic pseudo-Riemannian space-time.


\begin{thebibliography}{99}
\bibitem{Elli:1984} G.F.R. Ellis, in \textit{General Relativity and
Gravitation}, eds. B. Bertotti, F. de Felici and A. Pascolini (Reidel,
Dordrecht, 1984), p. 215.

\bibitem{Kras:1996} A. Krasi\'{n}ski, \textit{Inhomogeneous Cosmological
Models } (Cambridge University Press, Cambridge, 1997).

\bibitem{Boer:1998} J.P. Boersma, \textit{Phys. Rev. D} \textbf{57}, 798 (1998).

\bibitem{SHT:2007} W.R. Stoeger, A. Helmi and D.F. Torres, \textit{Inter. J.
Mod. Phys D} \textbf{16}, 1001 (2007).

\bibitem{Para-Sing:2008} A. Paranjape and T.P. Singh, \textit{arXiv}:0801.1546 [astro-ph].

\bibitem{Tava-Zala:1996} R. Tavakol and R. Zalaletdinov, \textit{Found.
Phys. } \textbf{28}, 307 (1998).

\bibitem{Zala:1992} R.M. Zalaletdinov, {Gen. Rel. Grav.} \textbf{24}, 1015 (1992).

\bibitem{Zala:1993} R.M. Zalaletdinov, {Gen. Rel. Grav.} \textbf{25}, 673 (1993).

\bibitem{Zala:1997} R.M. Zalaletdinov, \textit{Bull. Astr. Soc. India}
\textbf{25}, 401 (1997); \textit{gr-qc}/9703016.

\bibitem{Marc-Zala:1997} M. Mars and R.M. Zalaletdinov, \textit{J. Math.
Phys.} \textbf{38}, 4741 (1997).

\bibitem{Zala:1996} R.M. Zalaletdinov, \textit{Gen. Rel. Grav.} \textbf{28},
953 (1996).

\bibitem{Zala:2003} R.M. Zalaletdinov, \textit{Ann. European Acad. Sci.}
344 (2003).

\bibitem{Zala:2007} R.M. Zalaletdinov, \textit{gr-qc}/0701116.

\bibitem{Lore:1916} H.A. Lorentz, \textit{The Theory of Electrons} (Teubner,
Leipzig, 1916).

\bibitem{deGr-Sutt:1972} S.T. de Groot and L.G. Suttorp, \textit{Foundations
of Electrodynamics} (North-Holland, Amsterdam, 1972).

\bibitem{MRZ:2000} G. Montani, R. Ruffini and R. Zalaletdinov, \textit{Nuovo
Cim.} \textbf{115B}, 1343 (2000).

\bibitem{MRZ:2003} G. Montani, R. Ruffini and R. Zalaletdinov, \textit{Class.
Quantum Grav.} \textbf{20}, 4195 (2003).

\bibitem{CPZ:2005} A.A. Coley, N. Pelavas, and R.M. Zalaletdinov,
\textit{Phys. Rev. Lett.} \textbf{95}, 151102 (2005).

\bibitem{Zala-vdHo:2008} R.M. Zalaletdinov and R. van den Hoogen, to be submitted.

\end{thebibliography}
\end{document}